# Weighted-Codon-Usage Based Phylogeny In Ectocarpales

**Smarajit Das [a], Jayprokas Chakrabarti [a, b, *], Zhumur Ghosh [a], Satyabrata Sahoo [a, c] and Bibekanand Mallick [a]**

(a) Computational Biology Group
Theory Department
Indian Association for the cultivation of Science
Calcutta 700032 **INDIA**

(b) Biogyan
BF 286, Salt Lake
Calcutta 700064 **INDIA**

(c) Raidighi College
South 24 Paraganas , West Bengal **INDIA**

*Author for Correspondence:

e-mail:tpjc@iacs.res.in

Also : biogyan@vsnl.net

Phone: 91-033-2473 4971 EX: 307

Fax: 91-033-2473 2805

## Abstract:

We analyse forty seven chloroplastid genes of the large subunit of RuBisCO, from the Algal order Ectocarpales , sourced from GenBank. Codon-usage weighted by the nucleotide base bias defines our score called the Codon-Impact-Parameter . This score is used to obtain phylogenetic relations amongst the 47 Ectocarpales . We compare our classification with the ones done earlier.

## 1. Introduction:

Algae have grown in importance all over the world. Today they provide nutrition to millions. In this communication we study the family relations amongst an order of brown algae known as Ectocarpales. Brown algae are one of the major seaweeds. We analyse the chloroplastid gene , the large subunit of RuBisCO, to get the codon usage pattern. The score codon-impact-parameter is proposed . In the space of these scores the distance metric is defined and then used to estimate relations amongst Ectocarpales.

D-ribulose 1,5- bisphosphate carboxylase oxygenase , RuBisCO catalyses the carboxylation of ribulose 1,5 bisphosphate RuBP in the Calvin cycle . This bi functional enzyme at low carbon dioxide concentration and high oxygen level catalyses the oxygenation and cleavage of RuBP to form phosphoglycolate and 3-phosphoglycerate (Kellogg and Juliano, 1997). The large subunit of RuBisCO, denoted RbcL is co-transcribed with small subunit denoted RbcS. A spacer separates the two. Because of the nature of its specific function RbcL is a fairly conserved gene. The relatively low sequence divergence in RbcL and the importance of its function (Morton, 1993, 1996) have led to the assumption that change in codon usage is infrequent and unlikely for this gene.

Among the eukaryotic algae three major groups are recognized on the basis of photosynthetic pigmentation, viz Chlorophyta , Rhodophyta and Chromophyta . The Chromophyta is further subdivided into many small lineages. The brown algae or Phaeophyceae belong to this group. Ectocarpales are the most examined order within Phaeophyceae . RbcL has been studied widely for algae (Hommersand et al , 1994; Mc Court et al, 1995 ; Bailey and Freshwater, 1997 ; Daugberg and Anderson, 1997 ; Nozaki et al, 1997; Bailey et al 1998 ; Siemer et al 1998 ; Kogame et al, 1999; Hanyuda et al, 2000 ; Daugbjerg and Guilluo, 2000 ) .Ectocarpales have been chosen because of their simple structure ( Vanden Hoek et al, 1995 ). The cells of this family often have discoid chloroplast; the chlorophyll (a, $c_1$, $c_2$) is masked by other fucoxanthin pigment. The earlier classification of the Phaeophyceae was based on thallus construction, mode of growth and type of life history. In the early studies (Kylin, 1933; Vanden Hoek and Flinterman, 1968; Henry, 1984) the following orders were recognized in Phaeophyceae (i) Ascoseirales ( 1 family) , (ii) Cutleriales ( 1 family) , (iii) Desmarestiales ( 2 families) ( iv) Dictyotales ( 3 families) (De Clerck et al,2001; Lee and Bae, 2002) (v) Ectocarpales ( 23 families) , (vi) Fucales (about 8 families) (Rousseau and De Reviers, 1999) (vii) Laminariales ( 8 families) , (viii) Ralfsiales ( 2 families) , (ix) Scytothamnales ( 2 families) , (x) Sphacelariales ( about 4 families) , (x) Sporochnales ( 1 family) , (xii) Syringodermatales ( 1 family) and (xiii) Tilopteridales ( 1 family).

A clustering approach to brown algal classification using morphology, reproduction and ecological characters (Russell and Fletcher, 1975) was done earlier with 132 species. The significant outcome of this study was the suggestion to merge Chordariales , Dictyosiphonales and Tilopteridales and Scytosiphonales into Ectocarpales *Sensu lato* . Cytological characters , such as plastid structure, also play a major role for better understanding of the systematics of brown algae (Peters and Clayton, 1998; Muller et al, 1998) . Molecular phylogenies of the morphologically simple brown algae based on nuclear r-DNA sequences indicate that the taxa with predunculated pyrenoids form the monophyletic group the “Ectocarpales” ( Rousseau and De Reviers, 1999). Phylogenetic Relations (PR) based on molecular analysis in Chromophyta ( Bhattacharya et al, 1992; Leipe et al, 1994; Bhattacharya and Medlin, 1995; Medlin et al, 1997; Ben Ali et al, 2001; Draisma et al, 2001) have involved RuBisCO as well as 18s and 26s r-DNA. In recent years the molecular evidences using ribosomal genes and RuBisCO have divided Ectocarpales into five families namely, Chordariaceae , Actinetosporaceae , Adenocystaceae , Ectocarpaceae and Sytosiphonaceae

We study the codon-usage data of the large subunit of RuBisCO to define the Codon-Impact-Parameter (CIP). From the z-score of test statistic in the space of the CIP we identify with high confidence the most significant codons, the Impact Codons (IC). We find , these IC provide a sufficiently sensitive measure to explore phylogenetic relations, PR. We classify the relatively homogeneous order, Ectocarpales, belonging to brown algae (Phaeophyceae) phylum using the CIP scores of IC of RbcL. Using the standard pair group average analysis and the city block distance analysis we obtain PR among 47 Ectocarpales.

**2. Methods:**

Let PQR denote the codon. P is the nucleotide at the first codon position, Q the nucleotide in the second and R in the third. P, Q and R could be any of the four nucleotides A, C, G, T. The frequency, $F_{PQR}$, is given by

$$F_{PQR} = \frac{n_{PQR}}{\sum_{\alpha=1}^{20}\sum_{s=1}^{K_\alpha} X_{\alpha,s}} \qquad (1)$$

Where $n_{PQR}$ is the number of occurrences of the codon PQR. The index α is the amino acid index going from 1 to 20, and the symbol s refers to synonymous codons. The sum over s goes from 1 to $K_\alpha$, where $K_\alpha$ is the number of synonymous codons of the $\alpha^{th}$ amino acid. Thus $X_{\alpha, s}$ is the number of occurrence of the sth synonymous codon of the $\alpha^{th}$ amino acid.

The distribution of the nucleotides over the three-codon positions is different. The CIP score takes that into account. We define the CIP score of the codon PQR as:

$$(CIP)_{PQR} = \frac{F_{PQR}}{f_1(P) f_2(Q) f_3(R)} \qquad (2)$$

Where $f_1(P)$ is the frequency of the nucleotide P at the first codon position of the codon, $f_2(Q)$ is the frequency of the nucleotide Q at the second codon position and $f_3(R)$ refers to the frequency of the nucleotide R at the third codon position.

If $\overline{X}$ and μ denote the sample mean (mean of the CIP scores for a particular codon) and population mean (mean of sample mean) respectively; σ the population standard deviation, then z score of a test statistics is given by,

$$z = \frac{\overline{X} - \mu}{\sigma / \sqrt{N}} \quad \ldots\ldots\ldots \qquad \ldots(3)$$

Where N is the number of codons, i.e. 64.

. The PR in our case is derived as follows: If $\delta(c_l, c_m)$ is the distance between the Ectocarpales $c_l$ and $c_m$ ($l, m$ could take any value from 1 to 47, see Table1) then:

$$\delta(c_l, c_m) = \sum \left| CIP(c_l) - CIP(c_m) \right| \qquad (4)$$

Where the sum runs over all IC. The tree is generated by UPGMA method using arithmetic averages and city-block (Manhattan) distance analysis.

## 3. Discussions and Results :

The objective of the present study is to evaluate the codon usage of RbcL gene of Ectocarpales and reinvestigate PR within them to reassess the recently conceived family relations. For Ectocarpales about 47 RbcL sequences are available. These are chosen from GenBank and are denoted as $C_1$, $C_2$, ....$C_{47}$. ( Table 1)

For calculating the codon usage it is observed that some codons appear more often than expected from the underlying nucleotide composition of the gene. There are others that appear as expected and then there are others with lower than expected frequency (Ikemura, 1981, 1985; Gour and Gautier, 1982;

Bennetzen and Hall, 1982).The 'codon preference bias' proposed by Mclachlan et al (1984 ) quantify the degree of bias in codon usage and assess the relative merits of different codon from the viewpoint of translational efficiency. The 'Codon bias index' of Bennetzen and Hall (1982) calculate the frequency of optimal codons in a gene. Gribskov et al (1984) proposed another index the 'codon preference statistics' based on the ratio of the likelihood of finding a particular codon in a highly expressed gene to the likelihood of finding the codon in a random sequence with the same base composition.

Every model has its own applicative advantage over others. We feel that CIP score of a codon of a gene , as defined above, contain somewhat more quantitative information because it considers codon usage as well as the base compositional bias .We quantify the degree of codon bias in such a way that comparisons can be made both within and between species. Our approach to this problem is to define a measure for assessing the degree of relative abundance of a set of codons taking account of DNA modification and mutational rates on the compositional patterns of genomic sequences. Our statistic has the advantage that it measures the frequency of optimal codons from their underlying nucleotide composition of the gene. Moreover this codon bias statistic is defined in such a way that it has twin advantage of being simple to calculate yet making greater quantitative use of available information. We identify our impact codons, IC, chosen based on the level of significance from the z score of test statistic.

The ICs are selected based on the z score used in test statistic to establish and test hypothesis or the region of significance. If the null hypothesis that there is no significant difference is not rejected then there is no significant bias for the codon , and no over-representation of codons .One tailed test to examine the null hypothesis ($H_0$) , that CIP value of every codon is close to unity (as in random sequences) is performed .The CIP scores of the codons of RbcL differentiate the codons into different levels of significance. Three sets of codons identified at three different significance levels (0.01, 0.05, 0.1) and the alternative hypothesis, that a set of codons has a higher abundance with respect to their nucleotide composition, is accepted. These set of codons are IC. ICs are the ones that are significant above 90% level.

Classification of organisms based on molecular genomic methods depend crucially on the underlying arrangement / distribution of nucleotides or codons. If the biological group being studied is homogeneous, or the gene being studied for PR is well conserved (Kimura, 1981) the resolution of the individual entities from its nearest neighbors becomes ambiguous. Usually several alternate methods have to be used to map out the evolutionary pathway . The most common approaches to the tree construction involve UPGMA ( Sokal and Michener, 1958), neighbour-joining (Saitou and Nei, 1987), maximum parsimony (Farris, 1970) and maximum likelihood ( Cavalli-Sforza and Edward, 1967). A true biological phylogeny has a root or ultimate ancestor of all the sequences. Some algorithms like UPGMA provide information, or at least a conjecture, about the location of the root. Others like parsimony and the probabilistic models are uninformative about its position, and other criteria have to be used for rooting the tree.

We focus here on the distance method for tree building. Both the neighbour joining and UPGMA method use the distance-based algorithms but neighbour joining, unlike UPGMA, produces unrooted trees. The distance-based algorithms of tree building begin with a set of distances between each pair of sequences in a given dataset. There are many different ways of defining distance : Euclidean distance, City-Block distance, Chebychev distance, Power distance etc. . The Chebychev distance measure may be appropriate in cases when one wants to define two objects as 'different' if they are different on any one of the dimensions. Sometimes one may want to increase or decrease the progressive weight that is placed on dimensions on which the respective objects are very different. This can be accomplished via the Power distance. On the other hand the City-Block (Manhattan) distance is simply the average difference across dimensions. In most cases this distance measure yields results similar to simple Euclidean distance. This City-Block distance is computed from raw data . The advantage is that the distance between any two species is not affected by the addition of new species into the analysis. It needs to be emphasized that the different codon bias scores or the alignment algorithms, which estimate the differences between heterogeneous genes, fail for almost homologous genes. Here RbcL is well conserved both at amino acid and at codon level. For instance, some impact codons like Met ATG and Trp TGG are absolutely conserved within RbcL gene. But CIP values of these codons are different for different species due to unequal base composition. The CIP scores of IC of RbcL give us 47 matrices corresponding to 47 Ectocarpales (Table: 2) . Using the un-weighted pair group average analysis and the city block (Manhattan) distance analysis we obtain the PR amongst the 47 samples. We compare these with the presently conjectured relations ( Draisma et al, 2002). Interestingly, we find the matrices of CIP scores of just the IC largely reproduce the classification for the Ectocarpales . The few differences that we have are discussed.

The RbcL gene in all 47 cases is of length 1467 bp . Table 2 gives the CIP scores (Eqn 2) for the impact codons of RbcL. The z-score (Eqn 3) of the CIP-score-sample-statistic is significant; the null hypothesis rejected at 0.01 levels and IC are obtained. Thus at the 99% level of confidence ( Table 3 ) we have the following IC ; Tyr TAC, Phe TTC, Met ATG, Trp TGG, Leu TTA, Arg CGT, Asn AAC, Gly GGT, Ile ATC, Lys AAA, Gln CAA, and Glu GAA. These are 12 in number.

If we relax the level of significance and allow for z to be 0.05 (instead of 0.01) we have to add to the above list one extra codon , namely Thr ACT . That takes IC number to 13 at 95% level of confidence. And, if we relax our selection criteria a notch lower, z-score (Eqn 3) set at 0.1, we get one more codon in our list, namely Pro CCA. We use Eqn 4 to measure the distance between the matrices and then apply UPGMA to generate Fig 1. Similarly for z at 0.05 , the 47 matrices all have 13 rows and one column , we get Fig (2) . Fig (3) is for z at 0.1 level of significance. Fig 4 is the bootstrap consensus dendogram . It has been drawn by UPGMA method using Mega3 software.

In the bootstrap consensus tree, Fig 4, the position of the species largely agree with the present classification, Fig 5, of Ectocarpales ( Peters and Ramirez 2001, Draisma et al, 2002) . Some differences between the previous classification and our molecular tree are there for the genera *Colpomena, Scytosiphon* and *Petalonia*. Species of these genera , specially $C_{23}$ and $C_{24}$ , do not clade with $C_{21}$ and $C_{22}$ in our molecular phylogeny (Fig 1,2 & 3) and are scattered in the overall consensus bootstrap tree. One reason is, the phylogenetic tree built with just codon usage in RbcL is not sufficient . Other genes have to be considered as well. With just RbcL the amino acid content and the IC configurations of $C_{24}$ are identical with $C_{32}$ . Hence these cluster together. A close relationship between *Scytosiphon*, *Petalonia* along with *Giraudia* and *Sorocarpus* is suggested by our data.

The *Delamerea*, *Dictyosiphon*, *Punctaria* and *Coleocladia* ($C_1$, $C_2$, $C_6$ and $C_9$) clade is highly supported in all analysis (Fig 1,2,3). It is noted that both *Delamereae* and *Punctaria* have hecatonematoid microthalli and true Phaeophycean hairs with basal sheaths. Our study suggests that *Delamereaceae* is closely related to *Punctariaceae* (Siemer et al, 1998) along with *Dictyosiphonaceae* and *Coleocladiaceae.* Consensus bootstrap tree supports our data that $C_1$, $C_2$ and $C_6$ are closely clustered , but diverge for $C_9$. This divergence is noticeable for *Dictyosiphonaceae* in Figure 5. One reason is that the whole set of RbCL genes of Ectocarpales were not available earlier (Draisma et al (2002) ). But the divergence does not destroy the family structure of Chordariaceae . Chordariaceae is understood here as the largest and most diverse clade within Ectocarpales. Chordariaceae includes several species of so far accepted families, such as *Hummia* , *Myriotrichia*, *Stictyasiphon*, *Striaria*,

Isthmoplea, Litosiphon, Elachista, Hecatonema, Steblonema etc. Our data however suggest that there is one exception. Myelophycus ($C_{39}$, $C_{40}$) clusters with ($C_{34}$),which belong to Scytosiphonaceae.

Acinetosporaceae contains taxa with discoid plastids , a filamentous to parenchymatous thallus structure. In our analysis ,at 99% level of confidence, we find this clade contains the genera Pogotrichum ($C_{12}$), Pilayella ($C_{20}$), Feldmannia ($C_{44}$) and Hincksia ($C_{46}$). Here in our analysis three of these ( $C_{12}$, $C_{20}$ and $C_{46}$ ) belong to the clade along with Isthmoplea($C_{10}$) and *Chordaria*($C_{36}$).$C_{44}$ is in a separate cluster. Whereas at 95% level of confidence $C_{12}$ , $C_{20}$ and $C_{44}$ cluster but $C_{46}$ is in a separate group. Therefore , Acinetosporaceae overlaps with Chordariaceae and Scytosiphonaceae . Our phylogenetic analysis suggests that a further analysis , perhaps with other genes , is necessary to segregate these two families.

Ectocarpus silliculosus ($C_{13}$), the sole member of Ectocarpaceae ( Draisma et al, 2002) shows a clear divergence from Chordariaceae. In our analysis it is closely related to Acinetosporaceae and the genera Colpomenia($C_{23}$, $C_{24}$),Rosenvinegea ($C_{32}$)and Chnoospora ($C_{33}$) of Scytosiphonaceae. This finds support in Fig 4. In Fig 5 we notice the distinction between these three families .

We conclude that IC from CIP scores have the potential to sketch the PR among closely related species. The potential of the phylogenetic pattern is reduced if we introduce non-impact codons [$z_{critical} < 1.28$]. Also it is to be noted that if we plot the dendogram using just the non-impact codons-i.e, the ones not in the set above – the PR bear little resemblance to what is known about classification in the Ectocarpales.

Table :1 The 47 Phaeophyceae with their Genbank accession numbers .

| Species | GenBank Accession Number | Species | GenBank Accession Number |
|---|---|---|---|
| *Delamarea attenuata* ($C_1$) | **AF055396** | *Scytosiphon tenellus* ($C_{25}$) | **AB022241** |
| *Dictyosiphon foeniculaceus* ($C_2$) | **AF055397** | *Scytosiphon gracilis* ($C_{26}$) | **AB022240** |
| *Giraudia sphacelarioides* ($C_3$) | **AF055399** | *Scytosiphon canaliculatus* ($C_{27}$) | **AB022239** |
| *Hummia onusta* ($C_4$) | **AF055402** | *Scytosiphon lomentaria* ($C_{28}$) | **AB022238** |
| *Myriotrichia clavaeformis* ($C_5$) | **AF055408** | *Petalonia binghamiae* ($C_{29}$) | **AB022244** |
| *Punctaria plantaginea* ($C_6$) | **AF055410** | *Petalonia fascia* ($C_{30}$) | **AB022243** |
| *Stictyosiphon soriferus* ($C_7$) | **AF055413** | *Petalonia zosterifolia* ($C_{31}$) | **AB022242** |
| *Striaria attenuata* ($C_8$) | **AF055415** | *Rosenvingea intricate* ($C_{32}$) | **AB022232** |
| *Coelocladia arctica* ($C_9$) | **AF055395** | *Chnoospora implexa* ($C_{33}$) | **AB022231** |
| *Isthmoplea sphaerophora* ($C_{10}$) | **AF055403** | *Hydroclathrus clathratus* ($C_{34}$) | **AB022233** |
| *Litosiphon pusillus* ($C_{11}$) | **AF055406** | *Asperococcus fistulosus*($C_{35}$) | **AY095321** |
| *Pogotrichum filiforme* ($C_{12}$) | **AF055409** | *Chordaria flagelliformis*($C_{36}$) | **AY095324** |
| *Ectocarpus siliculosus* ($C_{13}$) | **X52503** | *Hecatonema sp . 86* ($C_{37}$) | **AF207802** |
| *Elachista fucicola* ($C_{14}$) | **AF055398** | *Microspongium globosum* ($C_{38}$) | **AF207805** |
| *Hecatonema sp* ($C_{15}$) | **AF055401** | *Myelophycus simplex* ($C_{39}$) | **AY095320** |
| *Streblonema tenuissimu* ($C_{16}$) | **AF055414** | *Myelophycus cavum* ($C_{40}$) | **AY095319** |
| *Laminariocolax tomentosoide* ($C_{17}$) | **AF055404** | *Polytretus reinboldii* ($C_{41}$) | **AF207809** |
| *Myrionema strangulans* ($C_{18}$) | **AF055407** | *Punctaria latifolia* ($C_{42}$) | **AY095322** |
| *Sorocarpus micromorus* ($C_{19}$) | **AF055411** | *Protectocarpus speciosus* ($C_{43}$) | **AF207810** |
| *Pilayella littoralis* ($C_{20}$) | **X55372** | *Feldmannia irregularis* ($C_{44}$) | **AF207800** |
| *Colpomenia phaeodactyla* ($C_{21}$) | **AB022237** | *Halothrix lumbricalis* ($C_{45}$) | **AF207801** |
| *Colpomenia bullosa* ($C_{22}$) | **AB022236** | *Hincksia hincksiae* ($C_{46}$) | **AF207803** |
| *Colpomenia peregrina* ($C_{23}$) | **AB022235** | *Mikrosyphar porphyrae* ($C_{47}$) | **AF207806** |
| *Colpomenia sinuosa* ($C_{24}$) | **AB022234** | | |

Table 2: The CIP Score of (12+1+1=14) Impact Codons (IC) of the Ectocarpales are listed.

First 12 Impact codons with 99% level of confidence. 13 th IC (Thr ACT) appears when the level of significance relaxed to 0.05. Pro CCA appears as 14 th IC when the significance level is 0.1.

| | Leu | Arg | Tyr | Phe | Met | Asn | Trp | Gly | Glu | Ile | Lys | Gln | Thr | Pro |
|---|---|---|---|---|---|---|---|---|---|---|---|---|---|---|
| | TTA | CGT | TAC | TTC | ATG | AAC | TGG | GGT | GAA | ATC | AAA | CAA | ACT | CCA |
| $C_{1}$ | 2.943 | 2.484 | 2.876 | 3.591 | 6.036 | 2.607 | 5.509 | 1.872 | 1.509 | 1.864 | 1.784 | 1.527 | 1.301 | 1.073 |
| $C_{2}$ | 2.942 | 2.336 | 3.121 | 3.478 | 6.085 | 2.311 | 5.563 | 1.944 | 1.484 | 1.416 | 1.843 | 1.544 | 1.328 | 1.266 |
| $C_{3}$ | 2.823 | 2.408 | 4.069 | 3.81 | 5.78 | 1.953 | 5.406 | 1.875 | 1.528 | 1.98 | 1.687 | 1.633 | 1.314 | 1.23 |
| $C_{4}$ | 2.67 | 2.419 | 3.647 | 3.026 | 5.318 | 3.126 | 5.265 | 1.93 | 1.493 | 1.321 | 1.909 | 1.47 | 1.33 | 1.208 |
| $C_{5}$ | 2.829 | 2.472 | 3.45 | 3.523 | 5.719 | 3.011 | 5.323 | 1.935 | 1.574 | 1.794 | 1.806 | 1.386 | 1.458 | 1.429 |
| $C_{6}$ | 2.925 | 2.505 | 3.195 | 2.937 | 5.919 | 2.889 | 5.359 | 1.831 | 1.5 | 1.879 | 1.787 | 1.517 | 1.322 | 1.066 |
| $C_{7}$ | 2.892 | 2.27 | 3.494 | 2.658 | 5.388 | 2.796 | 5.015 | 1.763 | 1.517 | 1.418 | 1.629 | 1.534 | 1.422 | 1.428 |
| $C_{8}$ | 2.898 | 2.471 | 3.763 | 3.524 | 6.036 | 3.01 | 5.618 | 2.049 | 1.501 | 2.05 | 1.784 | 1.507 | 1.312 | 1.235 |
| $C_{9}$ | 2.867 | 2.314 | 3.16 | 3.847 | 5.88 | 2.708 | 5.623 | 1.853 | 1.542 | 1.997 | 1.794 | 1.561 | 1.35 | 1.282 |
| $C_{10}$ | 2.94 | 2.171 | 2.309 | 3.243 | 5.567 | 2.748 | 5.132 | 1.82 | 1.531 | 1.872 | 1.861 | 1.588 | 1.296 | 1.281 |
| $C_{11}$ | 2.972 | 2.571 | 3.526 | 3.602 | 5.617 | 3.111 | 5.135 | 2.092 | 1.547 | 1.956 | 1.635 | 1.527 | 1.508 | 1.252 |
| $C_{12}$ | 2.596 | 2.185 | 2.56 | 2.963 | 5.669 | 2.469 | 5.374 | 2.034 | 1.421 | 2.286 | 1.558 | 1.451 | 1.251 | 1.359 |
| $C_{13}$ | 2.791 | 2.263 | 1.469 | 4.069 | 5.265 | 1.842 | 5.303 | 1.763 | 1.275 | 2.783 | 1.903 | 1.637 | 1.237 | 1.202 |
| $C_{14}$ | 2.905 | 2.482 | 3.916 | 3.362 | 5.571 | 2.679 | 5.081 | 2.011 | 1.594 | 1.729 | 1.653 | 1.536 | 1.405 | 1.072 |
| $C_{15}$ | 3.11 | 2.461 | 2.985 | 3.364 | 5.432 | 2.459 | 4.858 | 1.885 | 1.568 | 2.217 | 1.768 | 1.609 | 1.285 | 1.309 |
| $C_{16}$ | 2.795 | 2.514 | 3.499 | 3.574 | 6.036 | 3.032 | 5.619 | 2.085 | 1.555 | 2.144 | 1.784 | 1.375 | 1.483 | 1.235 |
| $C_{17}$ | 2.69 | 2.151 | 2.679 | 3.737 | 5.3 | 2.626 | 5.034 | 1.925 | 1.477 | 2.13 | 1.83 | 1.386 | 1.372 | 1.23 |
| $C_{18}$ | 3.072 | 2.427 | 3.371 | 3.703 | 5.437 | 2.543 | 5.011 | 1.954 | 1.458 | 1.172 | 1.718 | 1.332 | 1.476 | 1.334 |
| $C_{19}$ | 2.887 | 2.537 | 3.526 | 3.303 | 5.874 | 1.662 | 5.413 | 1.96 | 1.511 | 2.183 | 1.795 | 1.537 | 1.329 | 1.259 |
| $C_{20}$ | 2.873 | 2.284 | 2.619 | 3.567 | 5.616 | 2.393 | 5.031 | 1.941 | 1.256 | 1.711 | 1.675 | 1.491 | 1.161 | 1.233 |
| $C_{21}$ | 2.941 | 2.22 | 3.621 | 3.699 | 4.899 | 1.583 | 4.552 | 1.954 | 1.547 | 1.886 | 1.731 | 1.423 | 1.162 | 1.283 |
| $C_{22}$ | 2.941 | 2.248 | 3.726 | 3.49 | 4.899 | 1.493 | 4.552 | 1.979 | 1.547 | 2.034 | 1.731 | 1.423 | 1.177 | 1.283 |
| $C_{23}$ | 2.99 | 2.266 | 2.473 | 3.425 | 5.3 | 1.748 | 4.833 | 1.946 | 1.469 | 2.515 | 1.654 | 1.58 | 1.255 | 1.266 |
| $C_{24}$ | 2.795 | 2.21 | 2.107 | 3.638 | 5.3 | 1.445 | 4.83 | 1.904 | 1.468 | 2.668 | 1.562 | 1.528 | 1.132 | 1.235 |
| $C_{25}$ | 2.745 | 2.247 | 3.386 | 3.459 | 5.527 | 1.466 | 5.186 | 2.005 | 1.482 | 1.747 | 1.658 | 1.345 | 1.342 | 1.213 |
| $C_{26}$ | 2.87 | 2.225 | 3.416 | 3.489 | 5.389 | 1.493 | 4.954 | 2.016 | 1.509 | 1.78 | 1.689 | 1.527 | 1.113 | 1.083 |
| $C_{27}$ | 3.008 | 2.206 | 3.254 | 3.574 | 5.389 | 1.408 | 5.003 | 1.932 | 1.58 | 2.127 | 1.693 | 1.608 | 1.301 | 1.288 |
| $C_{28}$ | 2.876 | 2.257 | 2.966 | 3.61 | 5.013 | 1.398 | 4.751 | 1.945 | 1.54 | 1.891 | 1.815 | 1.405 | 1.187 | 1.247 |
| $C_{29}$ | 2.807 | 2.362 | 3.091 | 3.444 | 4.667 | 1.618 | 4.507 | 2.006 | 1.43 | 1.888 | 1.857 | 1.372 | 1.099 | 1.238 |
| $C_{30}$ | 2.756 | 2.218 | 3.731 | 3.494 | 5.719 | 1.724 | 5.377 | 1.995 | 1.413 | 1.761 | 1.741 | 1.319 | 1.277 | 1.19 |
| $C_{31}$ | 2.818 | 2.239 | 3.482 | 3.233 | 5.826 | 1.776 | 5.413 | 1.971 | 1.482 | 1.555 | 1.738 | 1.5 | 1.245 | 1.229 |
| $C_{32}$ | 2.899 | 2.178 | 1.783 | 3.541 | 5.351 | 1.684 | 4.855 | 1.966 | 1.426 | 2.628 | 1.654 | 1.517 | 0.999 | 1.043 |
| $C_{33}$ | 3.002 | 2.274 | 1.585 | 3.463 | 5.831 | 1.567 | 5.197 | 1.822 | 1.382 | 1.815 | 1.848 | 1.415 | 1.043 | 1.249 |
| $C_{34}$ | 3.008 | 2.114 | 1.53 | 2.774 | 5.094 | 2.266 | 4.516 | 1.925 | 1.397 | 1.775 | 1.719 | 1.252 | 1.076 | 1.06 |
| $C_{35}$ | 2.967 | 2.567 | 3.326 | 3.706 | 5.73 | 3.207 | 5.229 | 2.071 | 1.622 | 2.268 | 1.716 | 1.561 | 1.326 | 1.28 |
| $C_{36}$ | 2.858 | 2.243 | 1.799 | 3.675 | 5.583 | 2.18 | 5.145 | 2.177 | 1.503 | 1.979 | 1.696 | 1.5 | 1.349 | 1.23 |
| $C_{37}$ | 2.812 | 2.175 | 3.364 | 3.436 | 5.443 | 3.197 | 5.118 | 1.898 | 1.495 | 2.722 | 1.615 | 1.48 | 1.547 | 1.214 |
| $C_{38}$ | 2.8 | 2.52 | 3.506 | 3.581 | 6.048 | 3.039 | 5.63 | 2.089 | 1.558 | 2.149 | 1.787 | 1.372 | 1.486 | 1.237 |
| $C_{39}$ | 3.1 | 2.299 | 2.508 | 2.827 | 4.536 | 1.364 | 4.324 | 1.89 | 1.528 | 1.384 | 1.806 | 1.363 | 1.063 | 1.368 |
| $C_{40}$ | 3.147 | 2.289 | 2.257 | 2.943 | 4.444 | 1.578 | 3.971 | 1.881 | 1.554 | 1.067 | 1.856 | 1.188 | 1.134 | 1.342 |
| $C_{41}$ | 2.8 | 2.361 | 3.733 | 3.813 | 6.404 | 2.527 | 5.961 | 2.08 | 1.558 | 1.877 | 1.787 | 1.372 | 1.406 | 1.237 |
| $C_{42}$ | 3.136 | 2.456 | 3.053 | 3.441 | 5.311 | 2.516 | 4.749 | 1.936 | 1.581 | 2.268 | 1.783 | 1.474 | 1.282 | 1.32 |
| $C_{43}$ | 2.846 | 2.156 | 3.09 | 3.857 | 5.311 | 2.991 | 4.993 | 1.939 | 1.46 | 1.944 | 1.635 | 1.498 | 1.534 | 1.228 |
| $C_{44}$ | 2.856 | 2.377 | 3.633 | 2.855 | 4.808 | 3.112 | 5.145 | 2.041 | 1.406 | 1.816 | 1.534 | 1.467 | 1.311 | 1.203 |
| $C_{45}$ | 2.8 | 2.475 | 4.461 | 3.645 | 6.048 | 2.855 | 5.63 | 2.033 | 1.512 | 1.215 | 1.787 | 1.489 | 1.406 | 1.221 |
| $C_{46}$ | 2.96 | 2.175 | 1.666 | 3.403 | 5.838 | 2.697 | 5.37 | 1.946 | 1.504 | 2.204 | 1.486 | 1.549 | 1.242 | 1.269 |
| $C_{47}$ | 2.875 | 2.487 | 3.298 | 3.368 | 6.221 | 2.907 | 5.733 | 2.016 | 1.512 | 1.732 | 1.787 | 1.509 | 1.394 | 1.237 |

**Table 3 : z- values for codons . IC are highlighted**

| CODON | Corresponding $\overline{X}$ Values | Z values | CODON | Corresponding $\overline{X}$ Values | Z Values |
|---|---|---|---|---|---|
| GCT | 1.20268 | 1.21132 | CGG | 0 | -7.36538 |
| GCA | 1.19424 | 1.15114 | ATT | 0.98822 | -0.31805 |
| GCC | 0.87116 | -1.15285 | ATA | 0.24232 | -5.63732 |
| GCG | 0.83103 | -1.43903 | <u>ATC</u> | 1.91191 | **6.26908** |
| <u>GGT</u> | 1.95517 | **6.57758** | TAT | 0.97882 | -0.38509 |
| GGA | 0.55458 | -3.41049 | <u>TAC</u> | 3.05117 | **14.39353** |
| GGC | 0.73613 | -2.11579 | TCT | 0.83832 | -1.38704 |
| GGG | 0.57971 | -3.23128 | AGT | 0.41407 | -4.41251 |
| CTT | 0.83107 | -1.43874 | TCA | 0.68229 | -2.49974 |
| <u>TTA</u> | 2.89406 | **13.27312** | AGC | 0.1678 | -6.16874 |
| CTA | 0.74133 | -2.07871 | TCC | 0.08189 | -6.78140 |
| TTG | 0.00772 | -7.31033 | TCG | 0.13728 | -6.38639 |
| CTC | 0.00846 | -7.30505 | <u>AAA</u> | 1.7352 | **5.00890** |
| CTG | 0.13397 | -6.41000 | AAG | 0.62251 | -2.92605 |
| GTT | 0.90499 | -0.91159 | TTT | 0.51028 | -3.72640 |
| GTA | 0.52789 | -3.60082 | <u>TTC</u> | 3.44738 | **17.21904** |
| GTC | 0.03873 | -7.08919 | CCT | 1.15579 | 0.87694 |
| GTG | 0.84426 | -1.34468 | <u>CCA</u> | 1.24198 | **1.49159** |
| <u>ACT</u> | 1.29806 | **1.89151** | CCC | 0.84826 | -1.31615 |
| ACA | 1.05971 | 0.19176 | CCG | 0.27211 | -5.42487 |
| ACC | 0.20436 | -5.90802 | <u>ATG</u> | 5.51917 | **31.99367** |
| ACG | 0.4669 | -4.03576 | AAT | 0.3442 | -4.91078 |
| GAT | 0.99533 | -0.26735 | <u>AAC</u> | 2.32 | **9.17931** |
| GAC | 0.50881 | -3.73689 | <u>CAA</u> | 1.46921 | **3.11204** |
| <u>GAA</u> | 1.49518 | **3.29724** | CAG | 0.73894 | -2.09575 |
| GAG | 0.24278 | -5.63404 | CAT | 0.78297 | -1.78176 |
| <u>CGT</u> | 2.33334 | **9.27444** | CAC | 0.84628 | -1.33027 |
| AGA | 0.92972 | -0.73524 | TGT | 0.86632 | -1.18736 |
| CGA | 0.76312 | -1.92332 | TGC | 0.01852 | -7.23331 |
| AGG | 0.32997 | -5.01226 | <u>TGG</u> | 5.12951 | **29.21487** |
| CGC | 0.15507 | -6.25953 | | | |

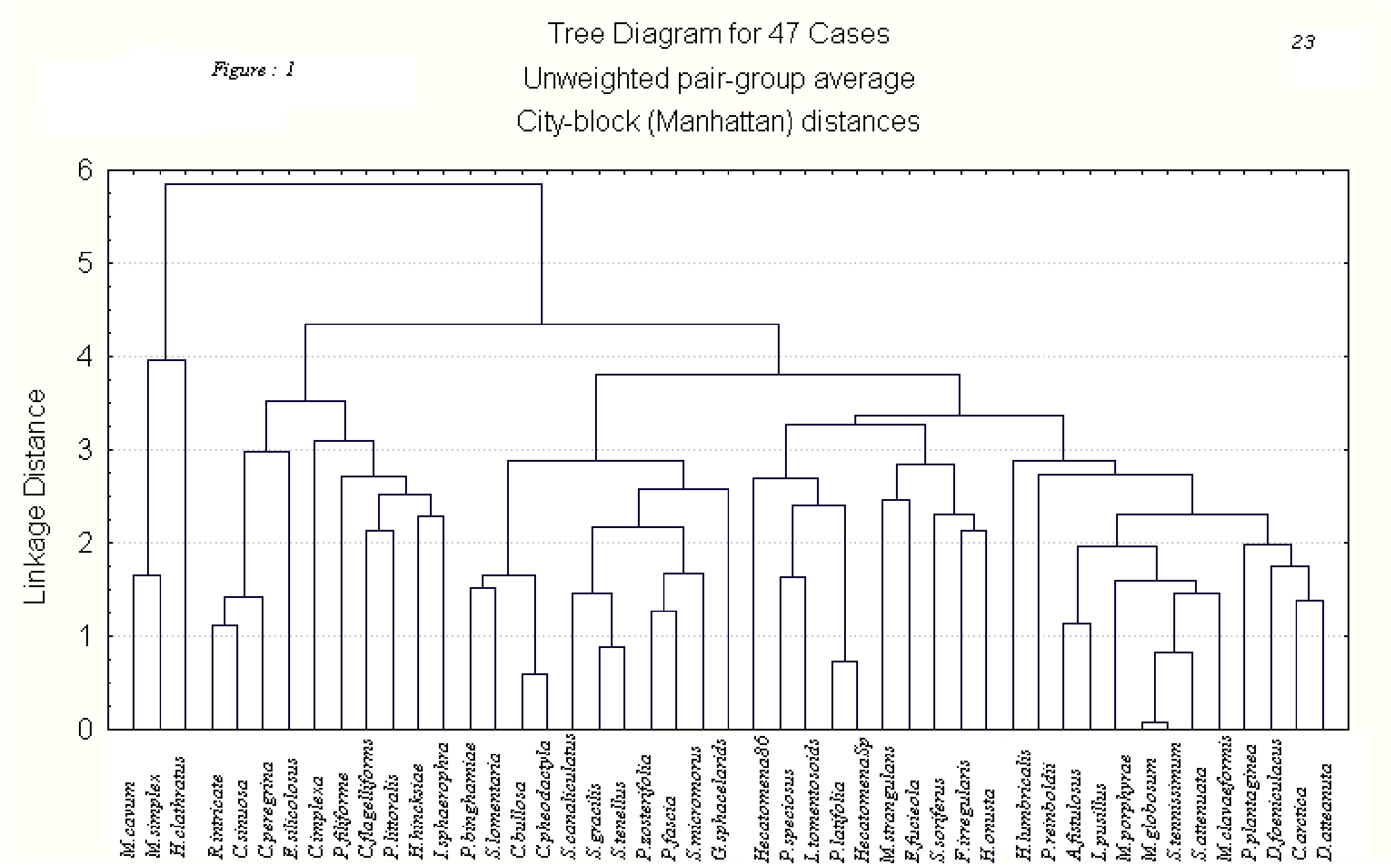

Phylogenetic analysis of 47 Ectocarpales species. The tree drawn according to Unweighted pair group method using arithmetic mean (UPGMA) and city-block (Manhattan) distance analysis at 0.01 level of significance. 12 codons have been used.

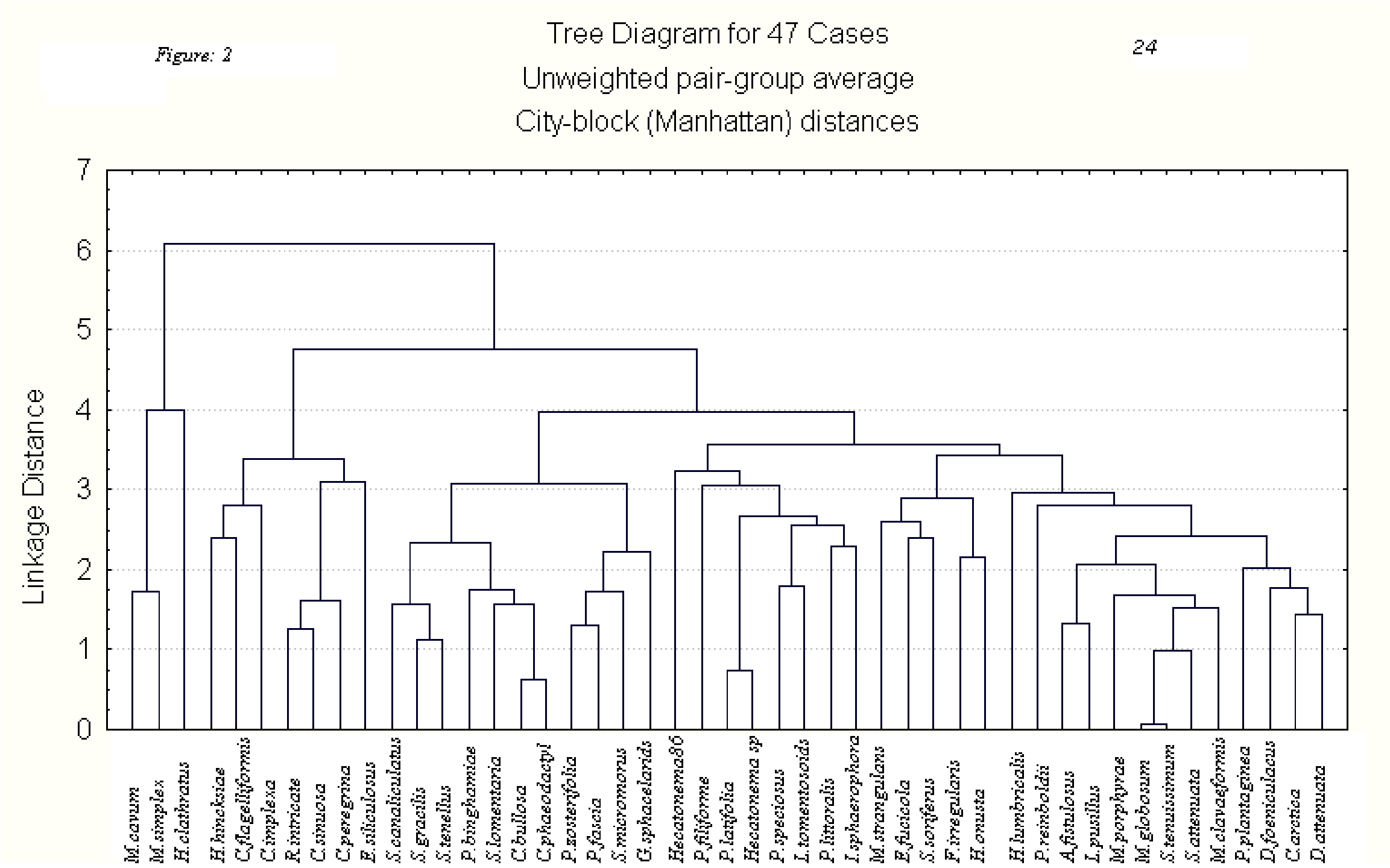

Figure 2: Phylogenetic analysis of 47 Ectocarpales species. The tree drawn according to Unweighted pair group method using arithmetic mean (UPGMA) and city-block (Manhattan) distance analysis at 0.05 level of significance. 13 codons have been used.

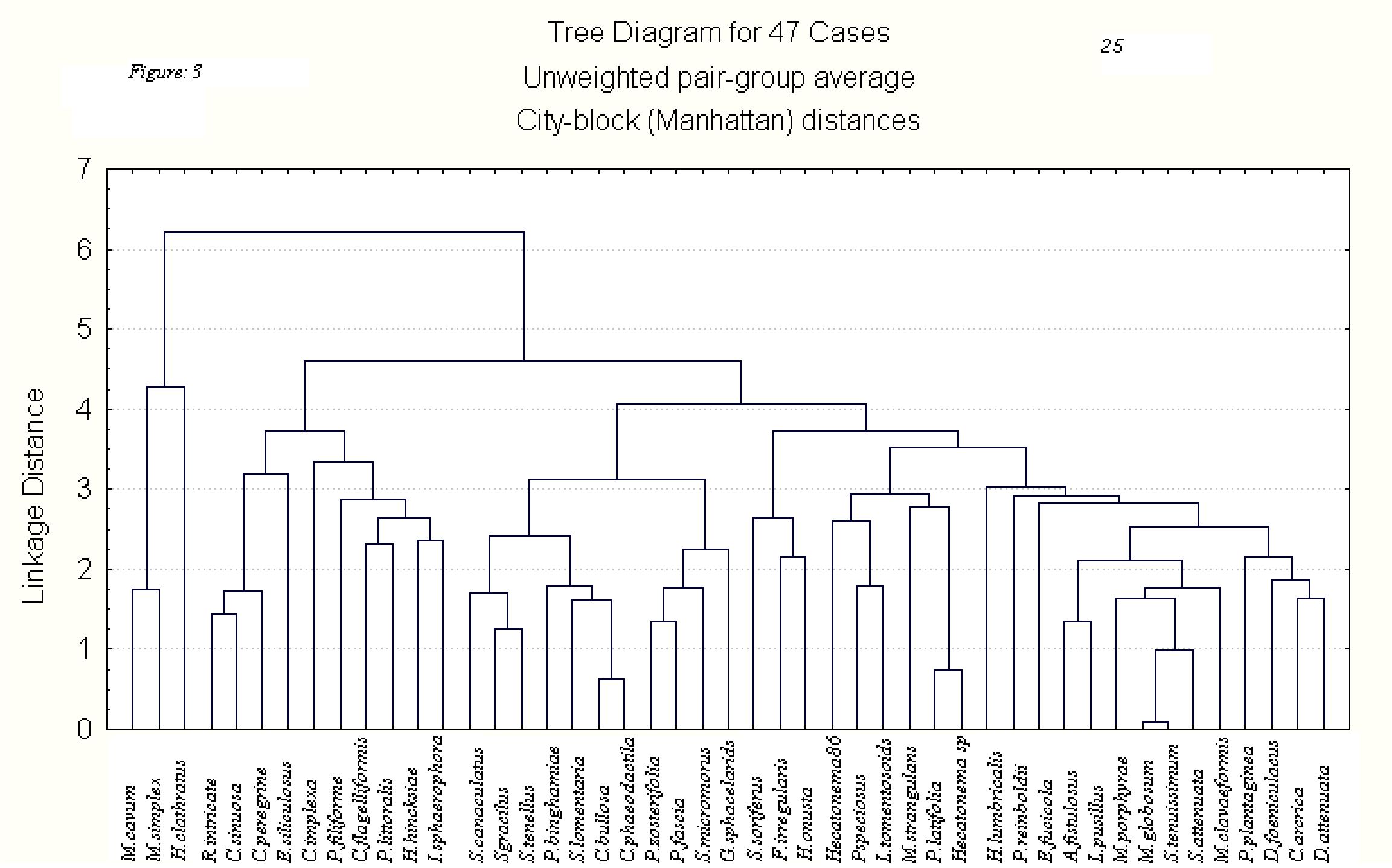

Figure 3: Phylogenetic analysis of 47 Ectocarpales species. The tree drawn according to Unweighted pair group method using arithmetic mean (UPGMA) and city-block (Manhattan) distance analysis at 0.1 level of significance . 14 codons have been used.

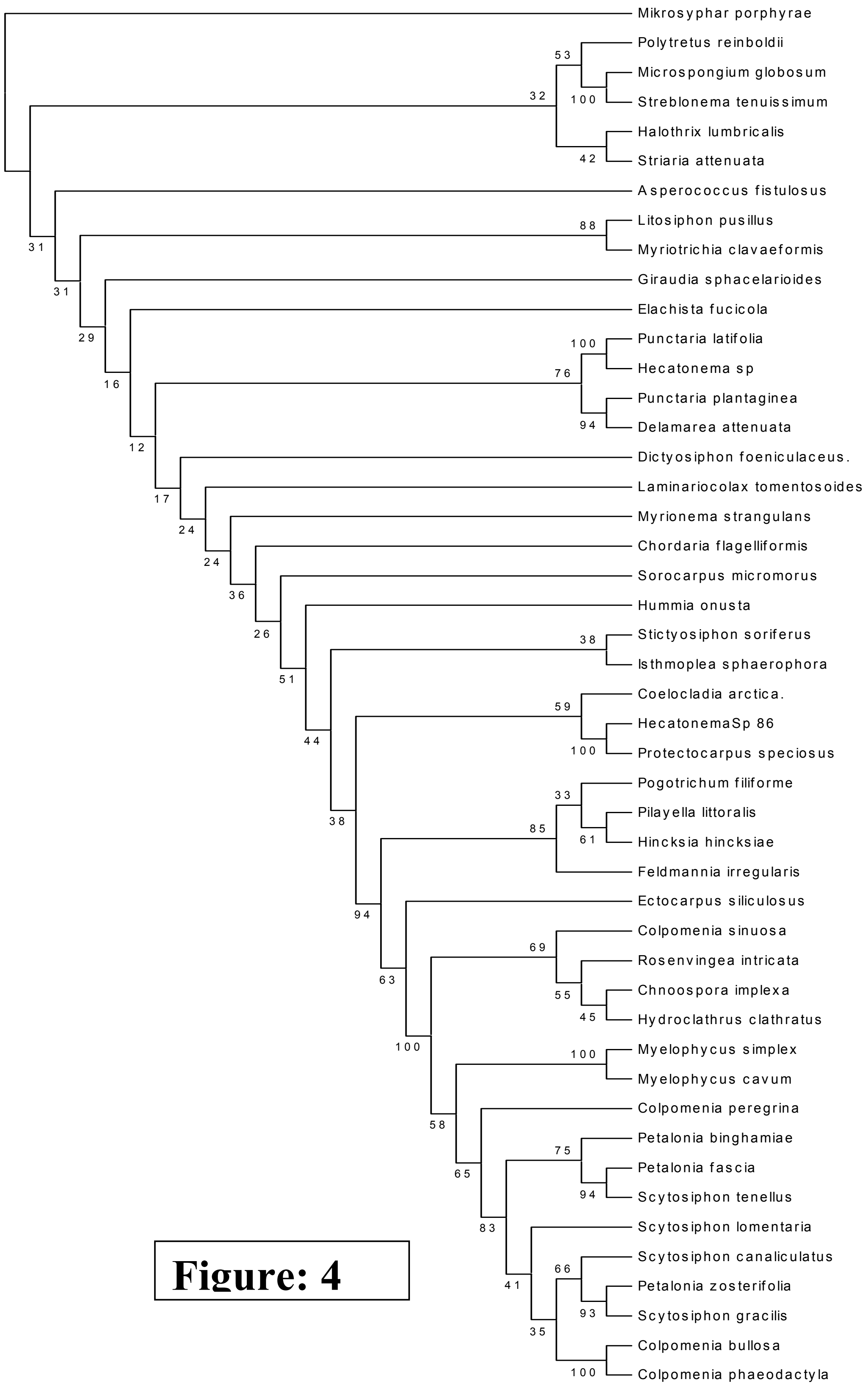

**Figure: 4**

Fig 4 Consensus bootstrap phylogeny of 47 Ectocarpales.

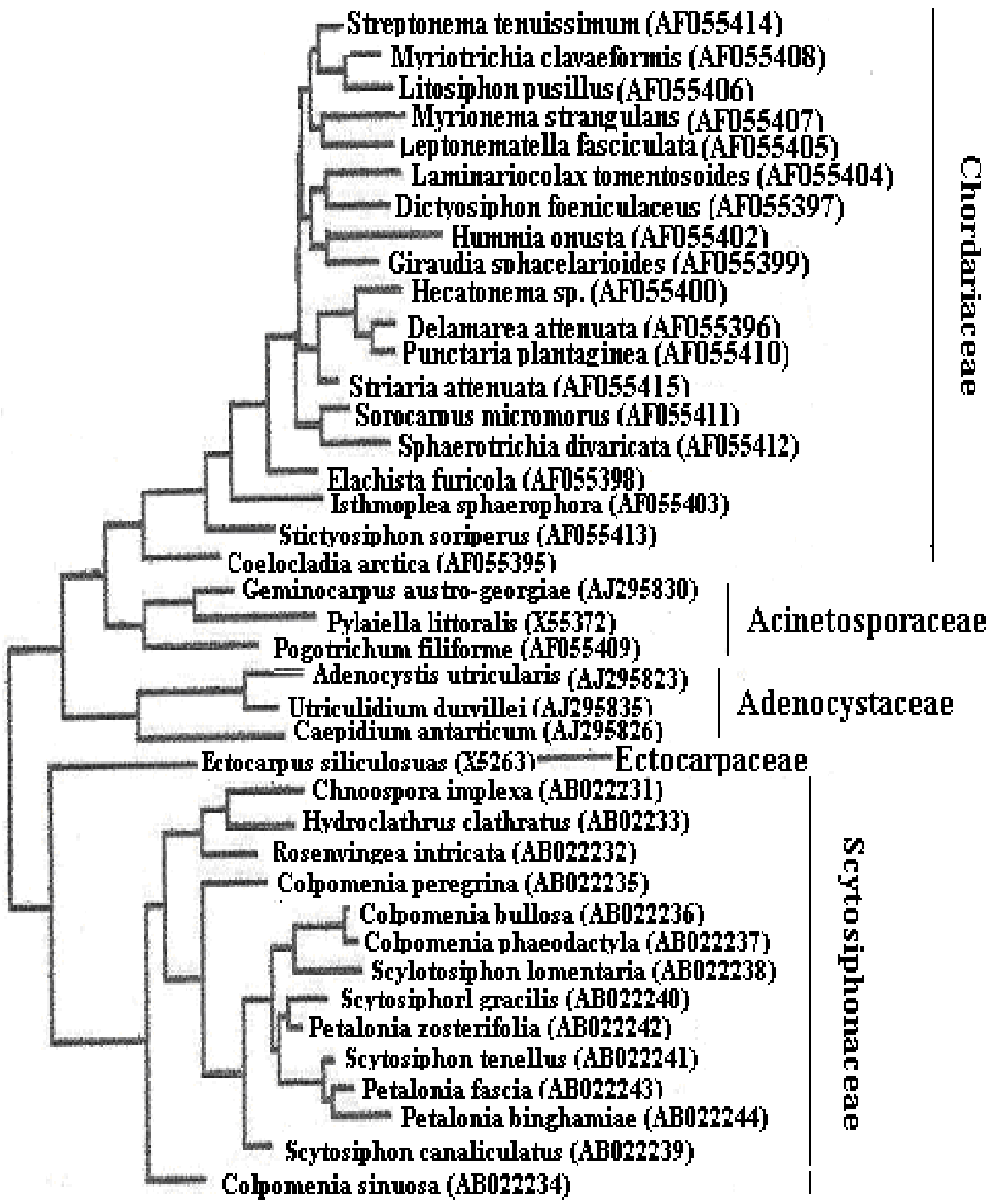

Figure 5

Figure 5: Present classification of Ectocarpales. See text for details.